\documentclass[preprint,12pt]{elsarticle}

\usepackage{graphicx}

\usepackage[table,xcdraw]{xcolor}

\usepackage{amssymb}
\usepackage{amsmath}
\usepackage{url}

\makeatletter
\def\ps@pprintTitle{%
 \let\@oddhead\@empty
 \let\@evenhead\@empty
 \def\@oddfoot{}%
 \let\@evenfoot\@oddfoot}
 
\makeatother

\begin{document}

\begin{frontmatter}

\title{Signatures of crypto-currency market decoupling from the Forex}

\author[ifj,pk]{Stanis{\l}aw Dro\.zd\.z}
\ead{stanislaw.drozdz@ifj.edu.pl}
\cortext[cor1]{Corresponding author}
\author[ifj]{Ludovico Minati}
\author[ifj]{Pawe{\l} O\'swi\c ecimka}
\author[pk]{Marek Stanuszek}
\author[ifj]{Marcin W\c atorek}

\address[ifj]{Complex Systems Theory Department, Institute of Nuclear Physics, Polish Academy of Sciences, ul.~Radzikowskiego 152, 31-342 Krak\'ow, Poland}
\address[pk]{Faculty of Physics, Mathematics and Computer Science, Cracow University of Technology, ul.~Warszawska 24, 31-155 Krak\'ow, Poland}

\begin{abstract}

Based on the high-frequency recordings from Kraken, a cryptocurrency exchange and professional trading platform that aims to bring Bitcoin and other cryptocurrencies into the mainstream, the multiscale cross-correlations involving the Bitcoin (BTC), Ethereum (ETH), Euro (EUR) and US dollar (USD) are studied over the period between July 1, 2016 and December 31, 2018. It is shown that the multiscaling characteristics of the exchange rate fluctuations related to the cryptocurrency market approach those of the Forex. This, in particular, applies to the BTC/ETH exchange rate, whose Hurst exponent by the end of 2018 started approaching the value of 0.5, which is characteristic of the mature world markets. Furthermore, the BTC/ETH direct exchange rate has already developed multifractality, which manifests itself via broad singularity spectra. A particularly significant result is that the measures applied for detecting cross-correlations between the dynamics of the BTC/ETH and EUR/USD exchange rates do not show any noticeable relationships. This may be taken as an indication that the cryptocurrency market has begun decoupling itself from the Forex.

\end{abstract}

\end{frontmatter}

\section{Introduction}
\label{Intro}
When Satoshi Nakamoto proposed the cryptocurrency Bitcoin (BTC) based on peer-to-peer network and encryption techniques~\cite{satoshi2008} in 2008, the blockchain technology was born. The idea behind this was providing, for the first time in human history, a tool thanks to which people anywhere could entrust each other and transact within an extensive network not requiring centralized management. The~methods on which the Bitcoin is based, as regards information storage, encryption technologies, and consensus protocols, were already established beforehand~\cite{watenchofer2016}. Nevertheless, as is often the case, for innovation to take place someone needs to combine existing technology in a new way and this must land on fertile ground, which was provided in 2009 by the aftermath of the financial crisis and resulted in the Bitcoin network as a distributed secure database. At that time, the Bitcoin quickly started getting wider recognition, not only within communities of tech geeks but also within the broader financial industry and, due to the anonymity of the transactions, even in the ``underworld'' of traders involved in dubious, when not outright illegal, businesses. The first fiat-to-bitcoin exchange, Mt.~Gox, was~launched in July 2010 and soon afterwards in February 2011 the first rules-free decentralized marketplace, called~Silk Road, where one could buy nearly any conceivable good using BTC, was~launched. These~events resulted in a drastically increased demand, leading to the first BTC bubble~\cite{sornette2018}, which~burst in the beginning of 2014 after the closure of Silk Road in October 2013 and the Mt.~Gox trading suspension in February 2014.\par
As the public awareness recognition of the Bitcoin increased, and more players developed an interest in the blockchain technology, it became apparent that the distributed ledger could be used not only as a basis for digital currencies but also for passing information and executing computer code on the blockchain. The idea of a globally-distributed cloud computing network, Ethereum, was proposed in late 2013 and then launched in July 2015. It allows anyone to create decentralized applications and own tokens by using smart contracts on the network. This capability provided the ground for the Initial Coin Offer (ICO) mania in 2017, which led to bubble engulfing the entire cryptocurrency market and eventually bursting in January 2018.\par
The current state of the blockchain technology could be compared to the dot-com bubble, which~unfolded at the turn of the last century. At that time, nearly everyone saw generic potential in internet technology, but it was not precisely known towards which direction the same would develop. In those times of euphoria, even rumors that a company started dealing with web technology would cause an increase in share price~\cite{Shiller2000}. Predictably, after the bubble burst, only a fraction of the leading companies survived.\par
{  Financial markets, especially the Forex market, due to their huge transactions volume, widely diversified participants and high speed of information processing, possess many of the emergent features that hallmark complex systems~\cite{kwapien2012}. A multitude of studies have analyzed the properties of the Forex market in terms of the returns distribution and volatility clustering~\cite{ausloos2000,drozdz2010}, persistence~\cite{tabak2006,berger2009}, multifractality~\cite{xu2003,drozdz2010} and cross-correlations~\cite{oh2012,gebarowski2019}. Recently, largely owing to its drastically higher volatility, the cryptocurrency market also gained research attention~\cite{kristoufek2013}. The studies published to date encompass market efficiency analysis~\cite{wei2018,kristoufek2019}, multifractal analysis~\cite{kristj2019,stosic2019} and cross-correlations analysis~\cite{stosic2018,zieba2019}; see Ref.~\cite{corbet2019} for additional references. However, mainly data with a temporal granularity (resolution) of one day have as yet been considered: evidently, this is inadequate given the high, and ever increasing, speed of information transmission. Here, the Forex market and the cryptocurrency market are compared, the latter represented at a more appropriate fine granularity of 10 s, as supported by the Kraken exchange data.}
\par
At the time of writing (May 2019), there are some 2200 active cryptocurrencies and tokens. New~blockchain-related projects and initiatives materialize at a remarkable rate, { (as exemplified by the Facebook coin (Libra)~\cite{libra}); applications in the energy sector related to the smart energy grid~\cite{pieroni2018}, aggregating multiple energy resources~\cite{tan2019}, and more broadly in the Internet of Things (IoT) receive increasing attention \citep{hang2019,nord2019}.} There is a clear (over)proliferation and fragmentation of cryptocurrencies, crypto-exchanges, and trading platforms~\cite{zunino2018}. Here, a speculation is put forward that the future may bring their closer integration, leading to the emergence of a marketplace more closely resembling, in terms of its statistical features, the established currency Forex market.\par

\section{Data specification and properties}
\label{data}
The data set used in the present study consists of the exchange rates reflecting the actual trade involving the Bitcoin (BTC), Ethereum (ETH), Euro (EUR) and US dollar (USD). On this basis, the following six exchange rates are defined: BTC/USD, BTC/EUR, ETH/USD, ETH/EUR, BTC/ETH, together with the EUR/USD, herein taken as a standard benchmark. Since 1 July  2016, all cryptocurrency transactions are tracked at a frequency of $\Delta t = 10$~s, with the resulting time-series being recorded by Kraken~\cite{Kraken}, which is the world's largest Euro-to-Bitcoin exchange. Operating since September 2013 as one of the longest-trading, continuously-running Bitcoin exchanges, it~has branches in Canada, the EU, and the US. Supported fiat pairs include the CAD, EUR, and~USD. Notable supported cryptocurrencies encompass the BTC, ETH, LTC, BCH, XRP, XMR, DASH, XLM, DOGE, EOS, ICN, GNO, MLN, REP, USDT, ZEC, ADA and QTUM, allowing both fiat-to-crypto and crypto-to-crypto trades. In the present study, the four most liquid pairs, namely BTC/EUR, BTC/USD, ETH/EUR, ETH/USD, and the most liquid crypto-to-crypto pair, BTC/ETH, are considered. The~Kraken API allowed seamless access to tick-by-tick data.

\begin{figure}[h!]
\centering 
\includegraphics[scale=0.5]{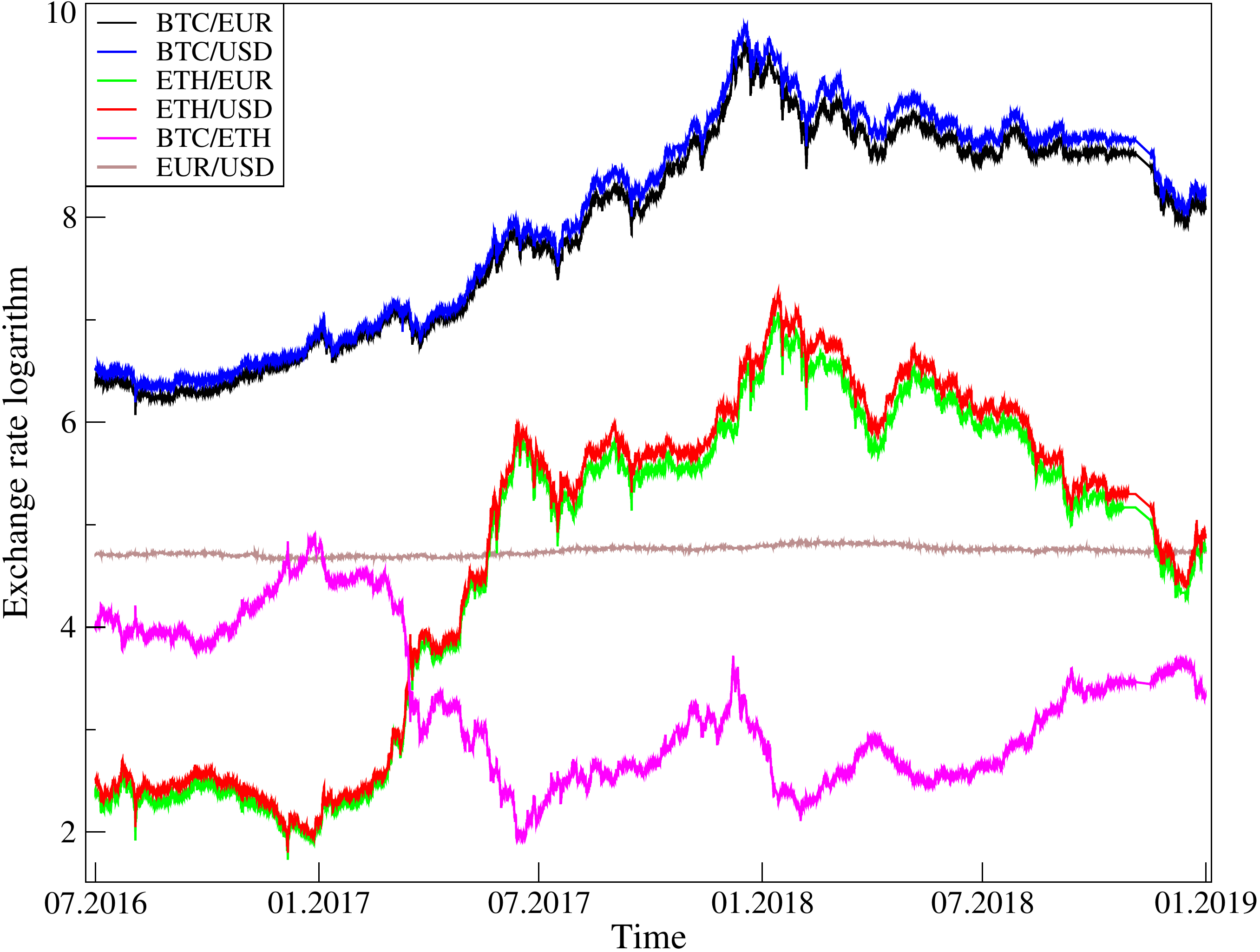}
\caption{(Color online) Logarithm of the exchange rates BTC/EUR, BTC/USD, ETH/EUR, ETH/USD, BTC/ETH and EUR/USD over the period between 1 July 2016 and 31 December 2018. For improved visibility, the EUR/USD exchange rate was magnified by a factor of 100.}
\label{fig:Exch}
\end{figure}

The data were collected until 31 December 2018. The EUR/USD exchange rate is considered at the same $\Delta t = 10$~s frequency, and within the same period of time but, due to Forex market trading specifications, without weekends (the Forex does not operate between Friday 10 p.m. UTC and Sunday 10~p.m. UTC), as recorded by the Swiss forex bank Dukascopy~\cite{dukascopy}. Charts~illustrating the time variation of these six exchange rates over the time-span under consideration are shown in Figure~\ref{fig:Exch}.\par

In the corresponding logarithmic returns, one has $r_{\Delta t}=\log(p(t+\Delta t))-\log(p(t))$, where $\Delta t$ stands for the returns' time-lag gaps, and where and the time intervals during which some instruments were not traded have been removed (7 May  2017 22.30--23.45---DDoS attack on ETH/USD, 6.40 11 January  2018--14.30 13  January  2018 Kraken maintenance shutdown). Thus, the series of returns from Kraken comprise approximately $N$ = 7.6 million observations for each of the considered time-series involving the BTC and ETH. For EUR/USD, there are about 5.6 million observations.

\begin{figure}[h!]
\centering 
\includegraphics[scale=0.15]{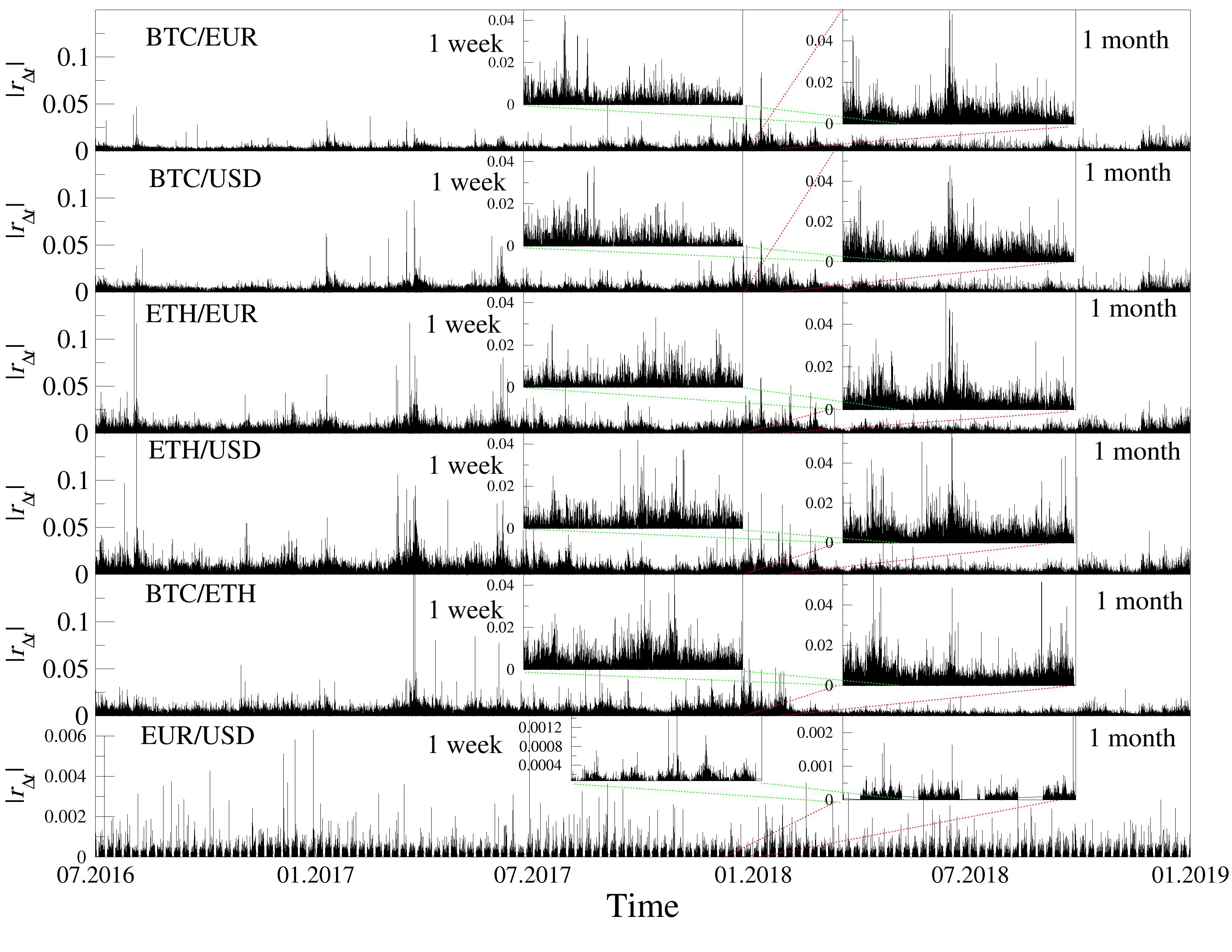}
\caption{(Color online) Time-variation of the moduli of $\Delta t = 10$~s logarithimic returns corresponding to the exchange rates of BTC/EUR, BTC/USD, ETH/EUR, ETH/USD, BTC/ETH and EUR/USD over the period between 1 July 2016 and 31 December 2018. The insets provide magnifications of the time-spans indicated.}
\label{fig:Volatility}
\end{figure}

Volatility clustering, a phenomenon reflecting the fact that large fluctuations tend to be followed by large fluctuations, of either sign, and small fluctuations tend to be followed by small fluctuations~\cite{mandelbrot1963}, is~considered to be among the most characteristic financial stylized facts~\cite{kwapien2012}. Such~an effect is clearly seen in Figure~\ref{fig:Volatility}, which shows the time-variation of the moduli of logarithmic returns for all six exchange rates under consideration. Remarkably, however, as demonstrated by the consecutive magnifications in the EUR/USD panel, the average time-span of the corresponding clusters in all the cases involving either the BTC or the ETH is about one order of magnitude longer than for the EUR/USD rate. In~the former case, one can recognize about three high-low volatility cycles within the corresponding monthly insets of Figure~\ref{fig:Volatility}, whereas~in the latter case, five such cycles are clearly apparent within the one-week time-interval of the inset in the bottom panel.

\begin{figure}[h!]
\centering 
\includegraphics[scale=0.5]{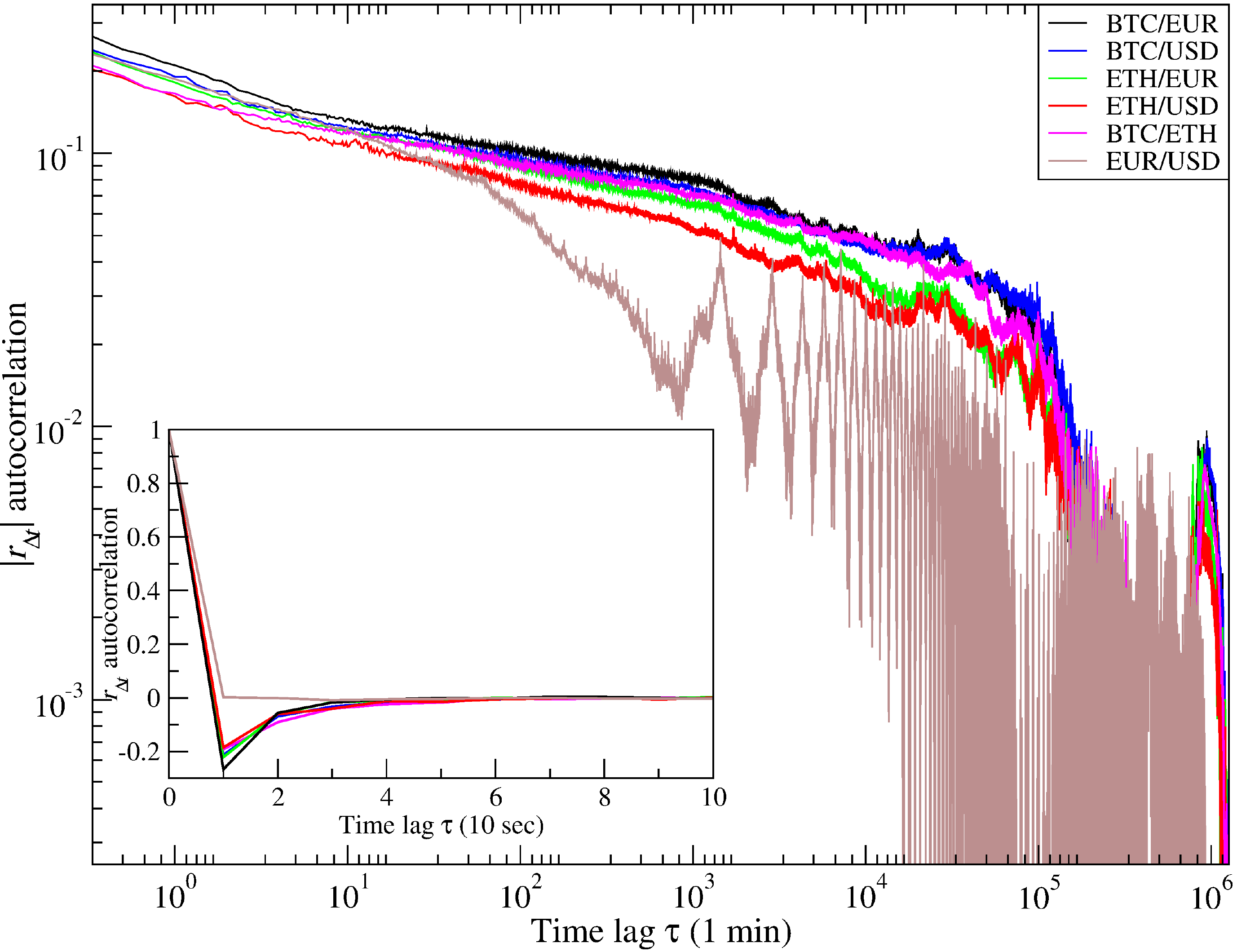}
\caption{(Color online) Volatility autocorrelation functions $C(\tau)$ for to the BTC/EUR, BTC/USD, ETH/EUR, ETH/USD, BTC/ETH and EUR/USD exchange rates over the period between 1 July 2016 and 31 December 2018. The daily trend was removed according to an established procedure~\cite{kwapien2012}.} 

\label{fig:volatility-auto}
\end{figure}

A more formal quantification of the corresponding effects can be arrived at via the volatility (logarithmic return modulus) autocorrelation function $C(\tau) = <\vert r(t)\vert \vert r(t-\tau) \vert >_t    $, as is shown in Figure~\ref{fig:volatility-auto} for all the six exchange rates under consideration. In order to suppress the strong seasonality inherent in financial dynamics, the daily trend was removed according to an established procedure~\cite{kwapien2012}, whereby at each time-point the signal is divided by the volatility mean standard deviation characteristic of that particular instant as evaluated from all the trading days included. Consistent decay following, to a good approximation, a power law of the form $C(\tau) \sim \exp(- \gamma)$ (straight line in the log-log scale) with $\gamma \approx 0.2$ is observed, but this kind of a decay clearly ends at about an order of magnitude lower $\tau$ for the EUR/USD exchange rate compared to the other exchange rates, involving either the BTC or the ETH. In the latter cases, this cut-off corresponds to about 10 days, whereas in the EUR/USD case, it corresponds to about one day. This truncation of the power-law scaling of $C(\tau)$, in general, reflects the average size of the volatility clusters in the time series of returns~\cite{drozdz2009}. In the present case, for the EUR/USD this appears to be about one order of magnitude faster than a decade ago on the same market~\cite{drozdz2010}, which tentatively agrees with the corresponding increased frequency of trading. Meanwhile, it appears noteworthy that it is the BTC and the ETH markets which, according to this analysis, develop dynamics resembling the EUR/USD one decade ago. Bearing in mind the remaining mismatch in the transaction capital involved, which clearly favors EUR/USD over the cryptocurrency market, such relative differences look remarkably consistent.\par     

One possible interpretation of this empirical fact is that the dynamics of cryptocurrency market, as measured in terms of the clock time, remains slower compared to the mature Forex currencies such as the EUR and USD. Equivalently, this would mean that the length of the volatility clusters is related more closely to the number of trades involved than to the physical clock time.\par

\begin{figure}[h!]
\centering 
\includegraphics[scale=0.5]{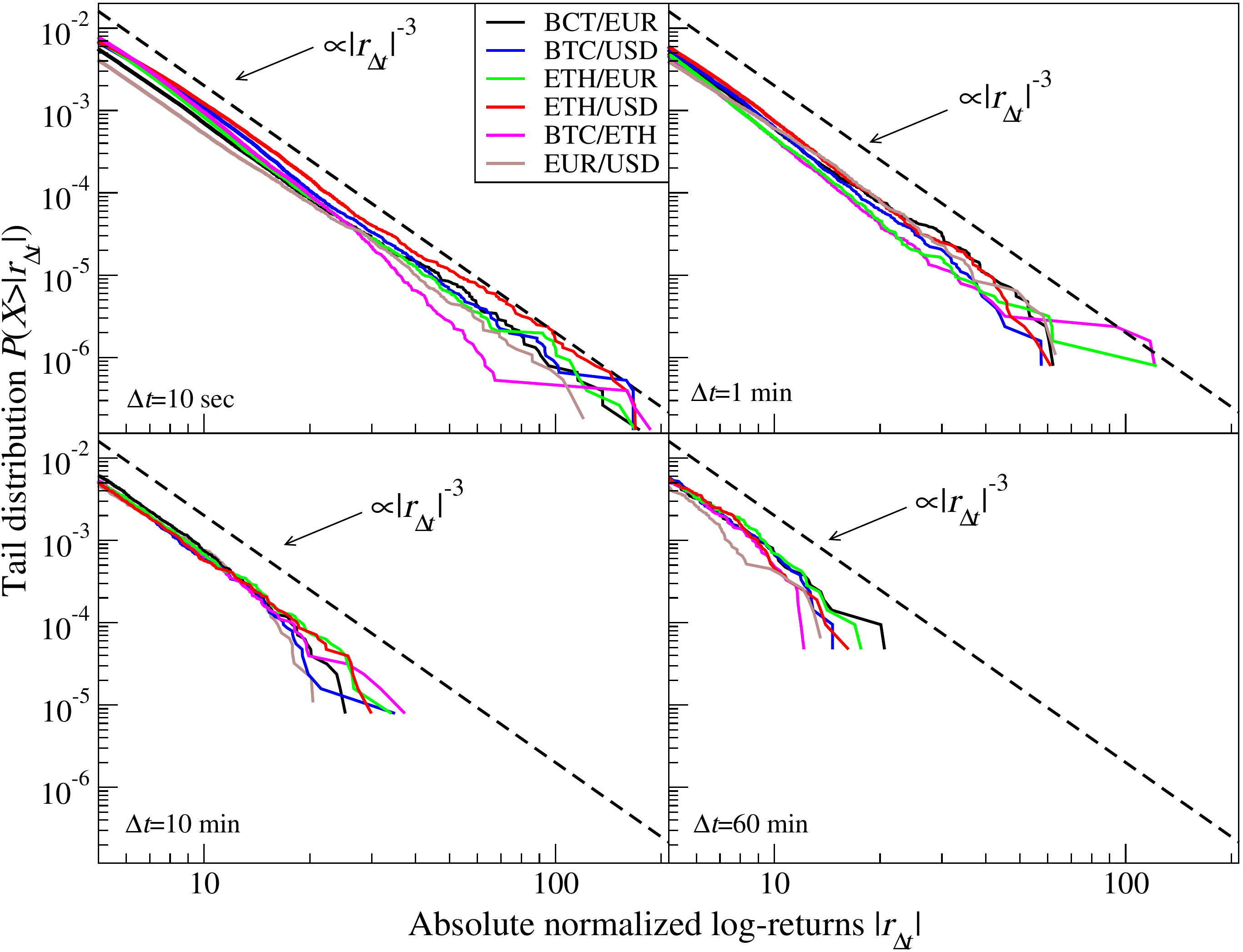}
\caption{(Color online) Log-log plot of the cumulative distributions of the normalized absolute returns $|r_{\Delta t}(t)|$ for BTC/EUR, BTC/USD, BTC/ETH, ETH/EUR, ETH/USD over the period between 1 July 2016 and 31 December 2018. The dashed line represents the expected inverse cubic power-law.}
\label{fig:distribution}
\end{figure}

The functional form of the statistical distribution $P_{\Delta t}(r)$ of the moduli of returns $\vert r_{\Delta t} \vert$ illuminates one of the most important characteristics of financial time series, with relevance also to multiscaling analysis. Namely, systematic studies~\cite{gopi1998,gopi1999,kwapien2012} of the stock market return distributions show that, on~sufficiently short time-scales $\Delta t$, the tails scale almost universally according to a power-law \mbox{$P_{\Delta t}(r) \sim r^{-\gamma}$}. For $P_{\Delta t}(r)$ taken in the cumulative form, this distribution asymptotically decays according to the inverse cubic power-law, i.e., $\gamma\approx 3$. Similar tail asymptotes are also evident for the conventional Forex market~\cite{drozdz2010}. In the older data from the capital market, this holds for $\Delta t$ up to several days but, in more recent data~\cite{drozdz2003,drozdz2007}, $P_{\Delta t}(r)$ is seen to bend downwards sooner (smaller $\Delta t$) towards a Gaussian distribution. The value of $\gamma$, then, becomes somewhat larger than $3$. This~effect may originate from the acceleration of information flow, together with a faster disappearance of correlations on larger time-scales when transitioning from the past to the present. In fact, the above discussion of the volatility of autocorrelation decay in the present EUR/USD rate in relation to its older behaviour~\cite{drozdz2010} provides a further indication of the validity of such a statement. As Figure~\ref{fig:distribution} shows for $\Delta t=$10~s, these distributions align remarkably well with the inverse cubic power-law for all the currency and crypto-currency pairs considered, including the BTC/ETH. With increasing $\Delta t$, which corresponds to aggregating smaller-scale fluctuations, the empirical distributions being systematically departing from this law, through bending downwards towards a Gaussian distribution; for $\Delta t =$ 60 min, this effect is already well-evident, as demonstrated in Figure~\ref{fig:distribution}. It is thus appropriate to assert that, insofar as the returns distribution is concerned, the exchange rates involving the two cryptocurrencies, BTC and ETH, obey the same law as the conventional currencies in the Forex.\par 

\section{Fundamental notions of the multifractal formalism}
\label{formalism}

The methodology applied herein for addressing the multifractality aspects of time-series is based on the formalism of multifractal cross-correlation analysis (MFCCA)~\cite{oswiecimka2014}. This method represents a consistent generalization of the detrended cross-correlation approach (DCCA)~\cite{podobnik2008} together with its multifractal extension~\cite{zhou2008}. The MFCCA methodology, in brief, allows quantifying both the scaling properties of individual time-series and the degree of multifractal cross-correlation between pairs of any two time-series. This methodology, then, introduced in a natural and self-consistent manner the $q$-dependent detrended cross-correlation coefficient $\rho_q$~\cite{kwapien2015}, such that the same allows filtering out the degree of strength of cross-correlations when it varies with the size of fluctuations.\par  

Let us consider a pair of time-series ${x(i)}_{i=1,...,T}$ and ${y(i)}_{i=1,...,T}$ divided into $2 M_s$ disjoint boxes having length $s$ (i.e., $M_s$ boxes starting from the opposite ends). The detrending procedure is then applied by calculating the residual signals $X,Y$ within each box $\nu$ ($\nu=0,...,2 M_s - 1$). These are equal to the difference between the integrated signals and the $m$th-order polynomials $P^{(m)}$ fitted to the same, namely
\begin{eqnarray}
X_{\nu}(s,i) = \sum_{j=1}^i x(\nu s + j) - P_{X,s,\nu}^{(m)}(j),\\
Y_{\nu}(s,i) = \sum_{j=1}^i y(\nu s + j) - P_{Y,s,\nu}^{(m)}(j).
\end{eqnarray}

In typical cases, an optimal choice is provided by $m=2$~\citep{oswiecimka2006,oswiecimka2013}, which is retained in the present~work.\par 

Next, the covariance and the variance of $X$ and $Y$ within a box $\nu$ are calculated according to
\begin{eqnarray}
\label{eq::covariance}
f_{XY}^2(s,\nu) = {1 \over s} \sum_{i=1}^s X_{\nu}(s,i) Y_{\nu}(s,i),\\
f_{ZZ}^2(s,\nu) = {1 \over s} \sum_{i=1}^s Z_{\nu}^2(s,i),
\label{eq::variance}
\end{eqnarray}
where $Z$ corresponds to either $X$ or $Y$. These quantities are subsequently used in defining a family of fluctuation functions having order $q$ \citep{oswiecimka2014}:

\begin{eqnarray}
\label{eq::covariance.q}
F_{XY}^q(s) = {1 \over 2 M_s} \sum_{\nu=0}^{2 M_s - 1} {\rm sign} \left[f_{XY}^2(s,\nu)\right] |f_{XY}^2(s,\nu)|^{q/2},\\
F_{ZZ}^q(s) = {1 \over 2 M_s} \sum_{\nu=0}^{2 M_s - 1} \left[f_{ZZ}^2(s,\nu)\right]^{q/2}.
\label{eq::variance.q}
\end{eqnarray}

The real-valued parameter $q$ plays the role of a filter, in that it amplifies or suppresses the intra-box variances and covariances such that, for large positive $q$-values, only the boxes (of size $s$) with the highest fluctuations contribute substantially to the sums in Equations~(\ref{eq::covariance.q}) and (\ref{eq::variance.q}), whereas, for negative $q$-values, only the boxes containing the smallest fluctuations provide a dominant contribution.\par 

The power-law dependence of $F_{XY}^{q}(s)$ on $s$ through the relation
\begin{equation}
F_{XY}^{q}(s)^{1/q}=F_{XY}(q,s) \sim s^{\lambda(q)},
\label{Fxy}
\end{equation}
is considered as a manifestation of the fractal character of the cross-correlations. Here, $\lambda(q)$ is an exponent that quantitatively characterizes the various fractality aspects. The $q$-independence of $\lambda(q)$, then, reflects monofractality of the cross-correlations. Contrariwise, the $q$-dependence of $\lambda(q)$ signals their more complex, multifractal character.\par

For a single time-series, possible scaling of the corresponding $F_{ZZ}$ reflects its scale-free properties
\begin{equation}
F_{ZZ}^{q}(s)^{1/q}=F_{ZZ}(q,s) \sim s^{h(q)},
\label{Fzz}
\end{equation}
where $h(q)$ stands for the generalized Hurst exponent. For multifractal signals, $h(q)$ is a decreasing function of $q$. The corresponding singularity spectrum $f(\alpha)$ can be determined using the following relations~\cite{kantelhardt02}:
\begin{equation}
\tau(q)=qh(q)-1,
\end{equation}
\begin{equation}
\alpha=\tau'(q)\quad \textrm{and} \quad f(\alpha)=q\alpha-\tau(q),
\label{falpha}
\end{equation}
where $\alpha$ is referred to as the singularity exponent or H\"older exponent, and $f(\alpha)$ is the corresponding singularity spectrum, often referred to as the multifractal spectrum. The particular case of individual time-series corresponds to the commonly used Multifractal Detrended Fluctuation Analysis (MFDFA)~\cite{kantelhardt02,kwapien2012,grech2016,jiang2018}. When $h(q)$ is approximately constant, the signal is interpreted as monofractal and $f(\alpha)$ collapses down to a single point.\par

The singularity spectrum $f(\alpha)$, corresponding to the moments (Equation~({\ref{eq::variance.q})) of order ranging between $-q$ and $+q$ for time-series generated by the model mathematical cascades, assumes the form of a symmetric upperward fragment of an inverted parabola. Realistic time series are, however, often distorted in their hierarchical organization as compared to a purely uniform organization of mathematical cascades~\cite{drozdz2015}. Such effects of non-uniformity typically manifest themselves in an asymmetry of $f(\alpha)$, and furthermore may also be crucially informative regarding the content of a given time-series. A straightforward approach to quantifying this kind of asymmetry of $f(\alpha)$ is through the asymmetry parameter~\cite{drozdz2015}:
\begin{equation}
A_{\alpha} = ({\Delta \alpha}_L - {\Delta \alpha}_R) / ({\Delta \alpha}_L + {\Delta \alpha}_R),
\label{ap}
\end{equation}
wherein ${\Delta \alpha}_L = {\alpha}_0 - {\alpha}_{min}$ and ${\Delta \alpha}_R = {\alpha}_{max} - {\alpha}_0$ and $\alpha_{min}$, $\alpha_{max}$, $\alpha_0$ denote the beginning and the end of $f(\alpha)$ support, and the $\alpha$ value at maximum of $f(\alpha)$ (which corresponds to $q=0$), respectively. The~positive value of $A _\alpha$ reflects the left-sided asymmetry of $f(\alpha)$, i.e., its left arm is stretched with respect to the right one. Since the left arm of $f(\alpha)$ is determined by the positive $q$-value moments, this~reveals a more developed multifractality at the level of large fluctuations in the time-series. Negative $A _\alpha$, on the other hand, reflects the right-sided asymmetry of the spectrum and thus illuminates a situation wherein it is the smaller fluctuations that develop richer multifractality.\par

The fluctuation functions defined by Equation~(\ref{Fxy}) can also be used to define a $q$-dependent detrended cross-correlation ($q$DCCA)~\cite{kwapien2015} coefficient:
\begin{equation}
\rho_q(s) = {F_{XY}^q(s) \over \sqrt{ F_{XX}^q(s) F_{YY}^q(s) }}.
\label{rhoq}
\end{equation}

This measure allows quantifying the degree of cross-correlations between two time-series $x(i)$ and $y(i)$, after detrending and at varying time-scales $s$. Furthermore, by varying the parameter $q$, one~can map out the size dependence of the strength of cross-correlations between the two signals. This filtering ability of $\rho_q(s)$ constitutes an important advantage over more conventional methods, since cross-correlations among the realistic time series are usually not uniformly distributed over their fluctuations of different magnitude~\citep{kwapien2017}. The $\rho_q(s)$ coefficient can, evidently, be applied in quantifying cross-correlations also between processes which develop no well established multifractality characteristics.\par

\section{Multifractality in the exchange rates}
\label{MFDFA}

Multifractality is a complex phenomenon arising and indexing several factors, in particular, long-range nonlinear temporal correlations, such as those revealed above, and the presence of heavy tails in the distribution of fluctuations. The first step towards multifractal analysis according to the commonly accepted procedure is to calculate the fluctuation functions as defined by Equation~(\ref{Fzz}), separately for each individual time series. In order to eliminate possible bias in estimating the fluctuation functions, the $q$-values span the range $q\in [-4,4]$. Due to the inverse cubic power-law~\mbox{\citep{gopi1998,drozdz2003}} governing the asymptotic distribution of large returns, also present herein as shown in Section~\ref{data}, this prevents entering the region of divergent moments for $q>4$. The range of scales, i.e., $s_{min}$ and $s_{max}$, for use in determining the scaling coefficients $h(q)$ depends in a first instance on the length $N$ of time-series under examination. A reliable estimation of the spectrum of $h(q)$ should take into account the fact that temporal correlations having long-range character are limited, as indicated in the previous section, to the average temporal size of the volatility clusters and, thus, do not encompass the entire span of the time-series: this, effectively, determines $s_{max}$. Clearly, all such clusters in the time-series are taken into account by the procedure, and it is in this sense that multifractality acquires its global characteristics. The other element, which determines $s_{min}$, is the length of occurrences of consecutive zero-valued returns. The presence of such sequences renders the negative $q$-value moments in Equation~(\ref{Fxy}) undefined.\par

The scale $s$-dependence of the fluctuation functions $F_q(s)$, determined by the Equation~(\ref{eq::variance.q}), for the collection of six exchange rates considered and for $q$-values ranging between $-$4 and 4 each is displayed in Figure~\ref{fig:qvariance}. The ones that correspond to the rates involving BTC or ETH, or both, develop undefined moments for negative $q$ within a range of small scales which reflect the presence of sequences of the consecutive zeros in these series. No such effects are seen for the EUR/USD exchange rate. Accordingly, also taking into account the signals coming from $C(\tau)$ of Figure~\ref{fig:volatility-auto}, the corresponding justified scale $s$ ranges between $s_{min}$ and $s_{max}$ are determined from the scaling coefficients $h(q)$ and are indicated by the vertical dashed lines. Insofar as scaling applies accurately, the precise determination of $s_{max}$ is not crucial, since varying it within reasonable limits does not influence the values of $h(q)$ in an appreciable manner. In the present case, the scale range is easily determined, and the corresponding values are shown in the corresponding panels of Figure~\ref{fig:qvariance}. On the other hand, for the EUR/USD pair, the same are almost symmetric with respect to $q=0$, whereas, for the exchange rates involving BTC or ETH, they develop a weaker or even no dependence on $q$ on the negative side. This, in fact, can even be seen visually from the relative locations of $f_{q}(\alpha)$ in this Figure. \par 

\begin{figure}[h!]
\centering 
\includegraphics[scale=0.47]{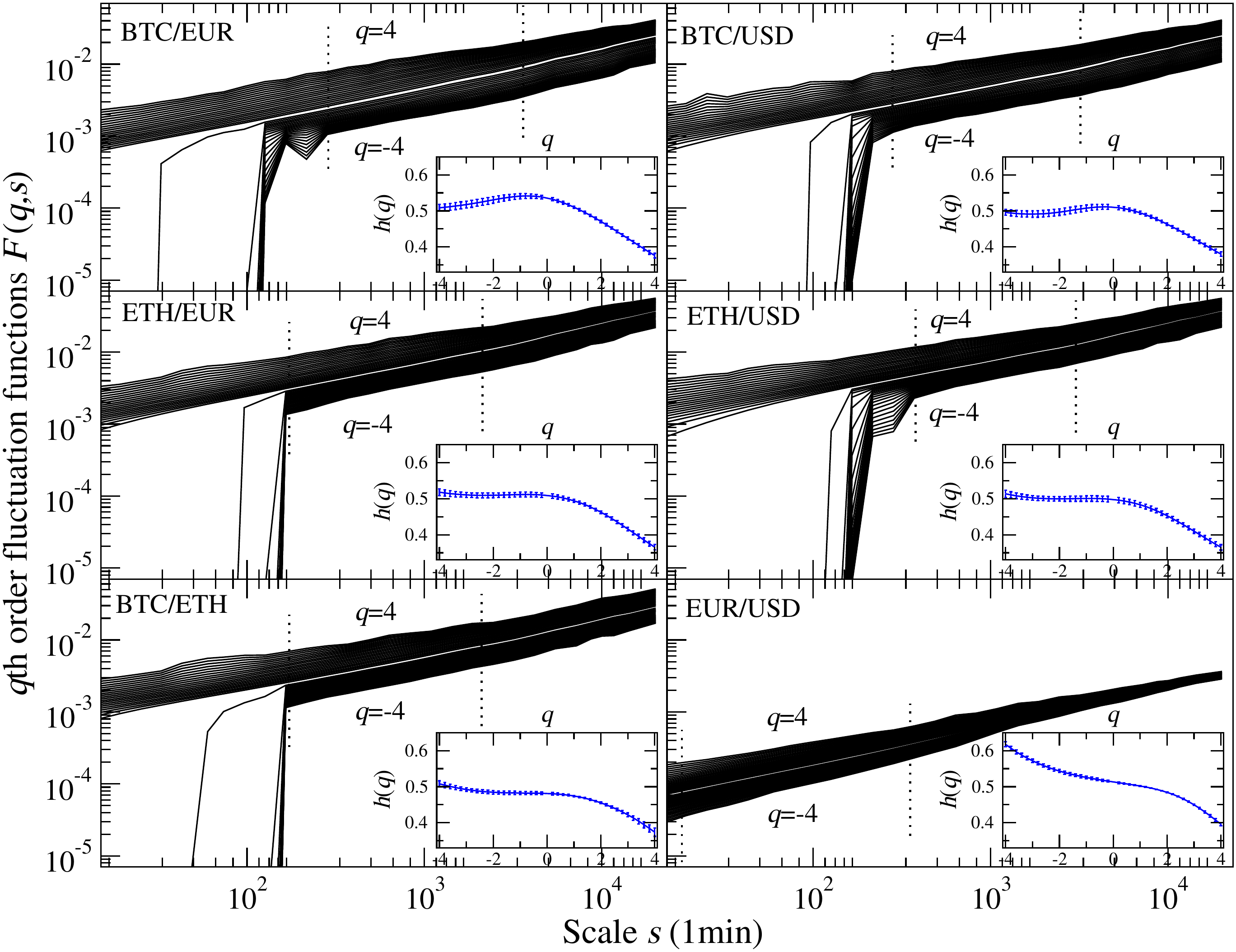}
\caption{Family of the $q$th-order fluctuation functions $F_q(s)$ for different values of $q\in [-4,4]$ in steps of 0.2 (the upper-most one represents $q=4$), calculated for the six exchange rates BTC/EUR, BTC/USD, ETH/EUR, ETH/USD, BTC/ETH and EUR/USD. Vertical dotted lines indicate the range of scales used in determining the generalized Hurst exponents $h(q)$. Insets show the corresponding dependence of $h(q)$ on $q$.}
\label{fig:qvariance}
\end{figure}

The differing multiscaling features are reflected in the shape of the multifractal spectra $f(\alpha)$ visible in Figure~\ref{fig:singspektrum}. Clearly, each of the considered time-series can be regarded as featuring a multifractal organization, manifest as a well-developed multifractal spectrum. However, with the exception of EUR/USD, whose $f(\alpha)$ spectrum in addition to being broad ($\Delta f(\alpha) \approx 0.75$) is almost symmetric $(A_{\alpha} \approx -0.02$), the other exchange rates develop evidently left-sided $(A_{\alpha}>0)$ multifractal spectra. Such~asymmetries of the multifractal spectra $f(\alpha)$, weather left- or right-sided, indicate non-uniformity~\cite{drozdz2015} in the hierarchical organization of the time-series. The left-side of $f(\alpha)$ is projected out by the positive $q$-values and as such reflects the organization of the large fluctuations, whereas the negative $q$-values determine the right-side of $f(\alpha)$. Consequently, the left-sided asymmetry of $f(\alpha)$ indicates a more pronounced multifractality at the level of large fluctuations, with the converse applying to a right-sided asymmetry. As Figure~\ref{fig:singspektrum} reveals, the left-sided asymmetry emerges in the exchange rates involving BTC and ETH which, in turn, indicates that they develop a more pronounced multifractality at the level of large fluctuations and that their smaller fluctuations are noisier. The most asymmetric cases are the BTC/EUR 
$(A_{\alpha} \approx 0.94)$ and the ETH/EUR $(A_{\alpha} \approx 0.92)$, where, effectively, only the left wing of the width $\Delta f(\alpha) \approx 0.35$ builds up. The BTC/USD $(A_{\alpha} \approx 0.72)$ and BTC/USD $(A_{\alpha} \approx 0.7)$ cases develop traces of a right wing in $f({\alpha})$, but the overall width, $(\Delta  f(\alpha) \approx 0.35)$, remains virtually unchanged. Notably, for BTC/ETH the right wing is even longer, however, the entire width extends to $\Delta f(\alpha) \approx 0.42$.\par

\begin{figure}[h!]
\centering 
\includegraphics[scale=0.48]{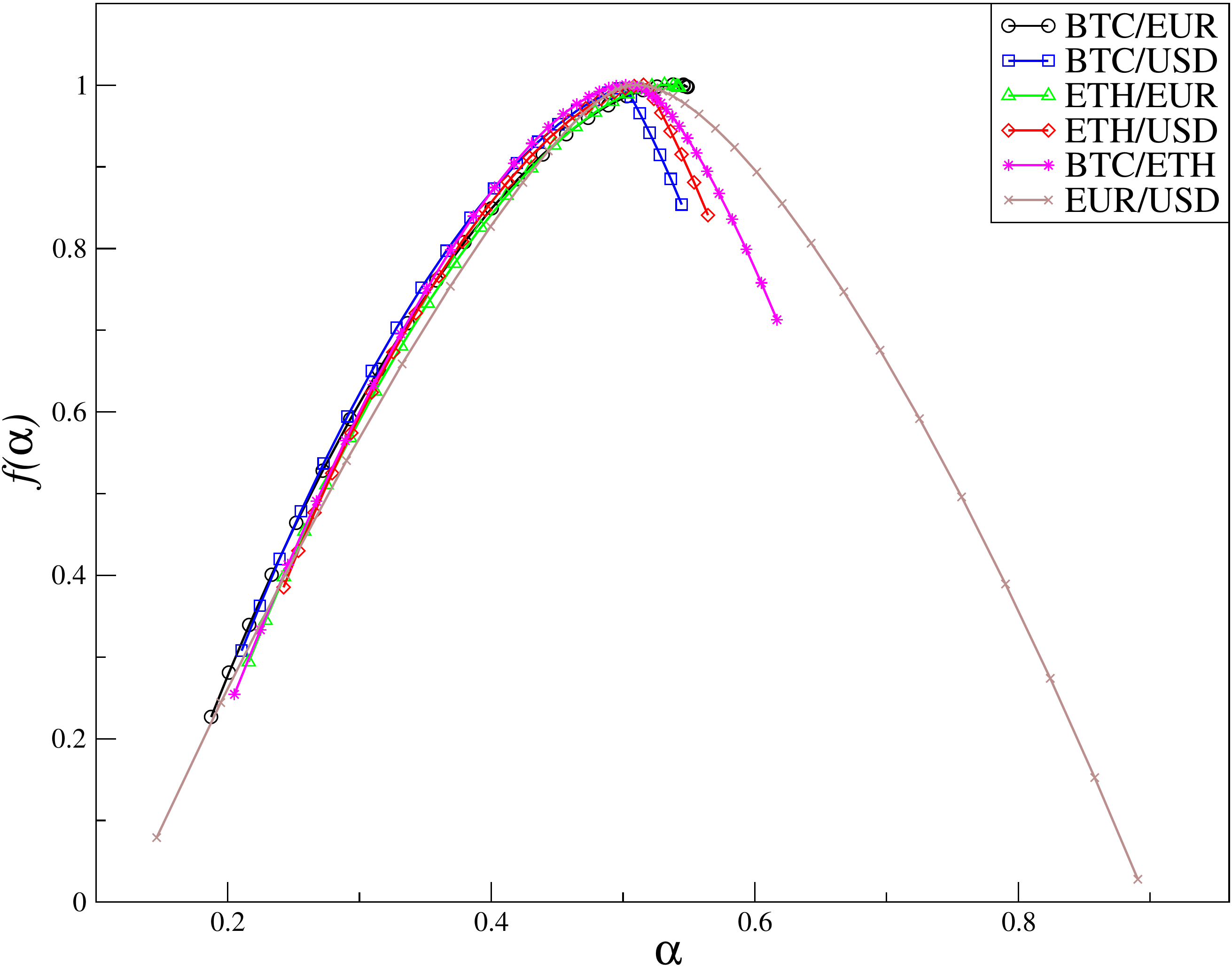}
\caption{(Color online) Singularity spectra $f(\alpha)$ calculated for all six considered time-series of BTC/EUR, BTC/USD, ETH/EUR, ETH/USD, BTC/ETH and EUR/USD returns, setting $q\in [-4,4]$. Insets show the $q$-dependence of the corresponding generalized Hurst exponents $h(q)$.} 
\label{fig:singspektrum}
\end{figure}

A commonly accepted and straightforward measure of correlations in time-series is provided by the Hurst exponent $H$~\cite{hurst1951}, which, within the present formalism, corresponds to $h(q=2)$. A finding of $H>0.5$ indicates persistence, whereas $H<0.5$ hallmarks anti-persistence. This parameter reveals the tendency to follow a trend and, as such, acquires particular relevance in the context of financial time series~\cite{ausloos2000}. In order to verify the anticipated evolution of such characteristics over time for the dynamics of the cryptocurrency exchange rates, the Hurst exponent was estimated as $H=h(q=2)$ following Equations~(\ref{eq::variance.q}) and (\ref{Fzz}) over 1-month (30 days) time windows, each comprising 259,200 data points (10~s returns). The results are visible in Figure~\ref{fig:Hurst}. The Hurst exponent $H(t)$ for the EUR/USD change rate was confirmed to dwell in the vicinity of 0.5, which hallmarks market maturity~\cite{matteo2003}. During the early trading period considered, i.e., between July 2016 and December 2017, the Hurst exponent $H(t)$ for the exchange rates BTC/EUR, BTC/USD as well as for ETH/EUR and ETH/USD had a value significantly smaller than 0.5: this revealed the strong anti-persistence which is expected for such a high-risk, emerging market. After that, the same parameter gradually approached 0.5 from below, eventually arriving very close to this value by the end of the year 2017. This can be interpreted as a strong indication that the corresponding markets are approaching maturity~\cite{matteo2003}, a fact that has recently been pointed out for BTC/USD rate~\cite{BTC2018}. At the same time, it is compelling to observe that the Hurst exponent for BTC/ETH exchange rate, while still markedly below 0.5 towards the end of the year 2017, started approaching this value later in the year 2018.\par
  
\begin{figure}[h!]
\centering 
\includegraphics[scale=0.5]{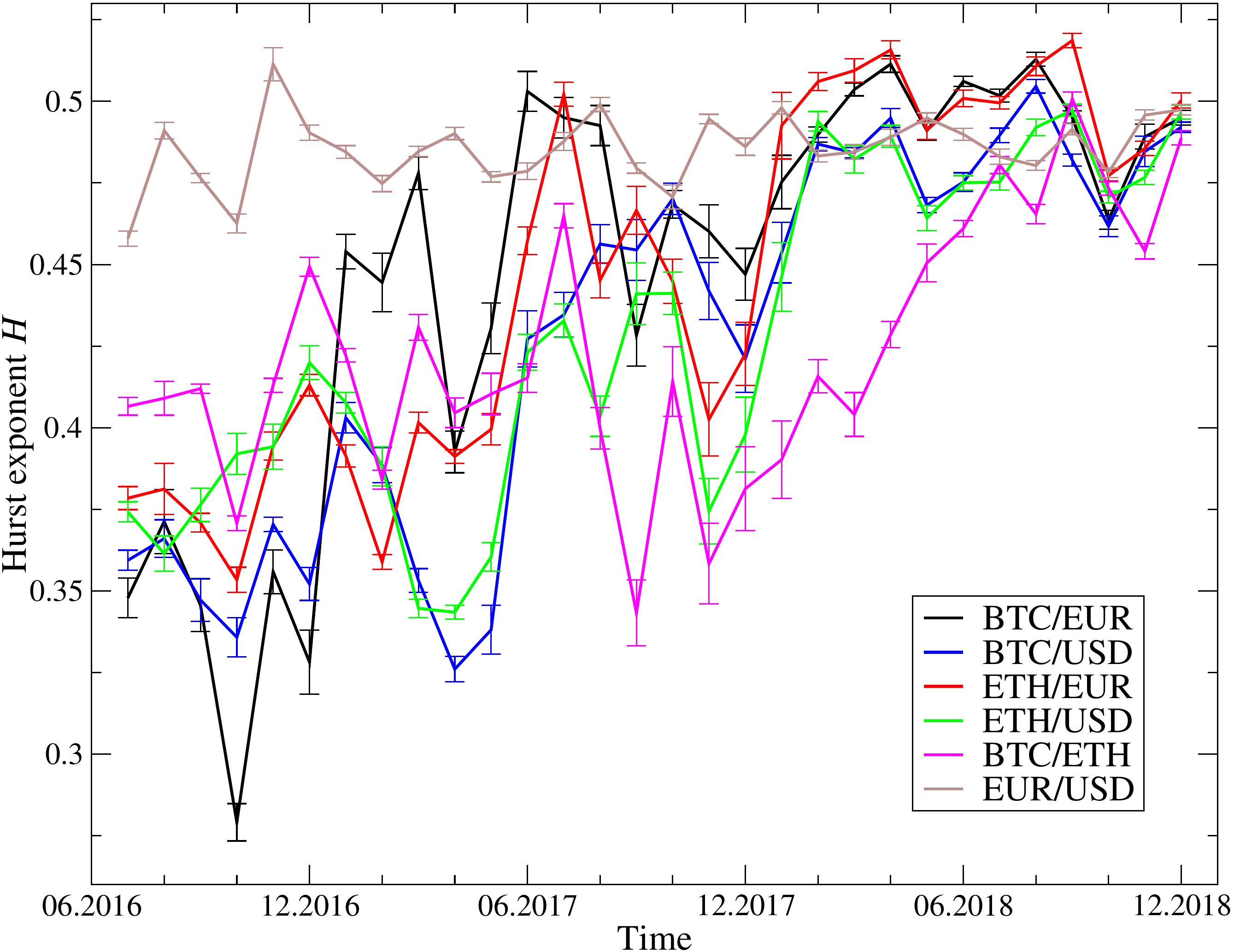}
\caption{Hurst exponent $H$ calculated over 30-days window for the BTC/EUR, BTC/USD, ETH/EUR, ETH/USD, BTC/ETH and EUR/USD exchange rates from  1 July  2016 to 31 December 2018. Error bars reflect the standard error of the regression slope.}
\label{fig:Hurst}
\end{figure}

\section{Cross-correlations and their mutliscaling features}
\label{MFCCA}

The currency exchange rates in the world foreign exchange market (Forex) delineate a rich landscape of cross-correlations, which are endowed with signatures of multiscaling~\cite{gebarowski2019}. In order to extend the present study towards this direction, i.e., towards the quantification of cross-correlations between the different pairs of exchange rates involving the BTC and/or the ETH, the corresponding $q$th-order fluctuation functions $F_{XY}(q,s)$ according to Equation~(\ref{Fxy}) were calculated, followed by searching for possible evidence of scaling. The results of such calculations are shown in Figure~\ref{fig:Fq-cc}, and reveal a rather convincing scaling behavior of $F_{XY}(q,s)$ for all cases under consideration. In~turn, this indicates that there exists a certain level of synchrony in the evolution of the corresponding exchange rates, even at the level of their multifractal organization. Power-law behavior of $F_{XY}(q,s)$ is, however, seen~exclusively for the positive $q$-values; therefore, these are shown in Figure~\ref{fig:Fq-cc} with the lower limiting values of $q$ listed explicitly. Below those values, the cross-correlation fluctuation functions $F_{XY}(q,s)$ start fluctuating irregularly, occasionally even becoming negative, as similarly observed in other financial phenomena considered previously~\cite{Rak2015,Watorek2019}. Within such limits, the most extended scaling, as far as the range of $q$-values is concerned, was detected for the BTC/EUR versus the BTC/USD, ETH/EUR or ETH/USD. Plausibly, this is likely related to the fact that changes in the EUR/USD exchange rate itself are predominantly on small scales, whereas the rates of BTC and ETH are considerably larger. Contrariwise, the least extended cross-correlation scaling was found for the BTC/ETH versus either the BTC/EUR or the BTC/USD.\par

\begin{figure}[h!]
\centering 
\includegraphics[scale=0.5]{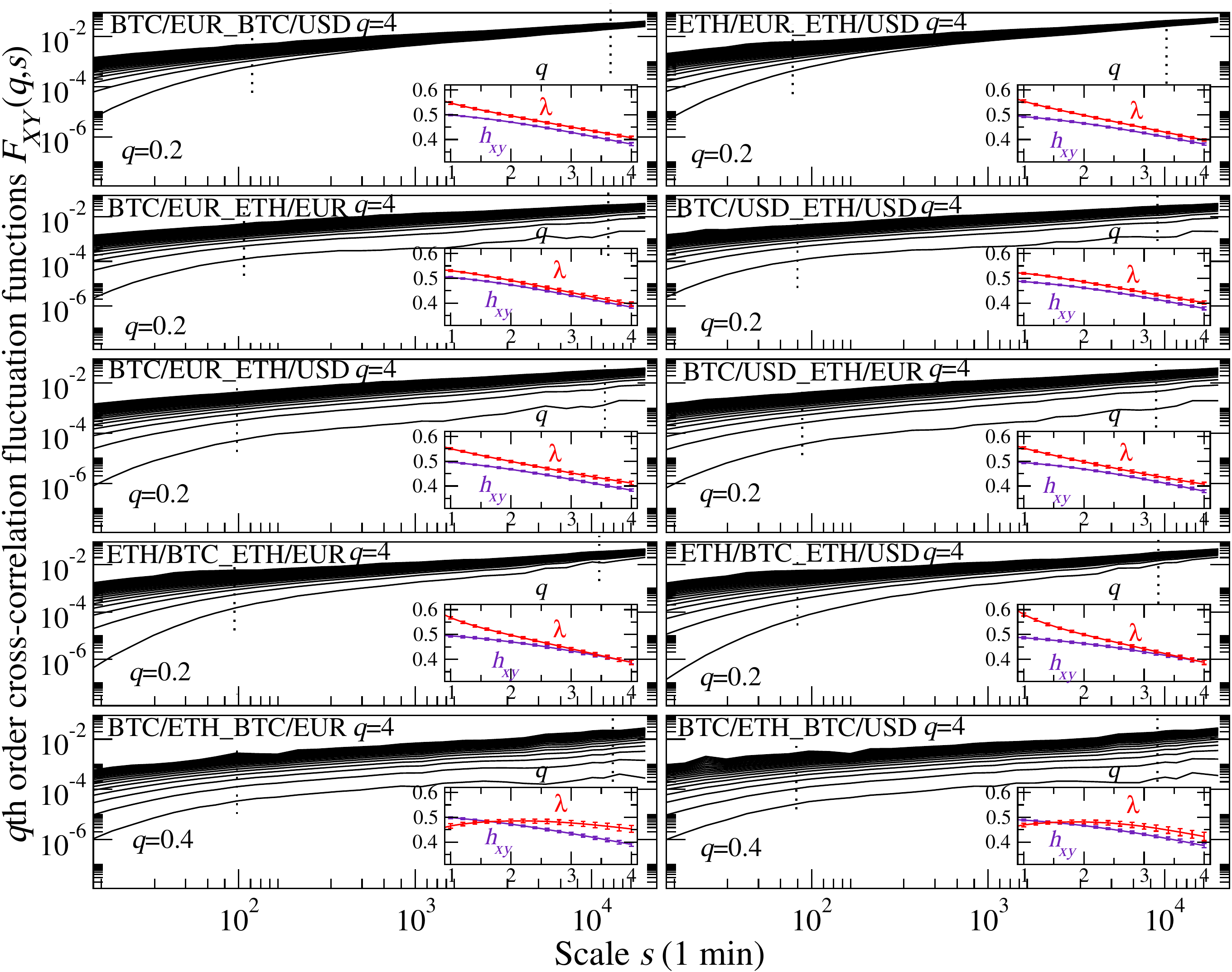}
\caption{Family of the $q$th-order fluctuation cross-covariance functions $F_{XY}(q,s)$, for different values of $q$ in steps of 0.2, and calculated for the cross-correlations among the BTC/EUR, BTC/USD, ETH/EUR, ETH/USD, BTC/ETH and EUR/USD exchange rates. The upper-most lines correspond to $q=4$, whereas the bottom ones to those for which $F_{XY}(q,s)$ are still positive. Insets illustrate the resulting $q$-dependence of $\lambda(q)$ versus the average of the generalized Hurst exponents $h_{xy}(q) = (h_x(q)+h_y(q))/2$ of the two series $x(i)$ and $y(i)$ under study.}
\label{fig:Fq-cc}
\end{figure}

Some further, more subtle effects related to the cross-correlations could be quantified through comparing the corresponding scaling exponents $\lambda(q)$ and the average generalized Hurst exponents $h_{xy}(q) = (h_x(q)+h_y(q))/2$. Here, $h_{xy}(q)$ behaved like all other cases considered, but this was accompanied by a noticeable variation in the $q$-dependences of $\lambda(q)$. This result, shown in Figure~\ref{fig:lambda_h-average} for correspondingly different $q$-dependences of $d_{xy}(q) = \lambda(q) - h_{xy}(q)$, effectively reflects a changing rate of covariance accumulation in Equation~(\ref {eq::covariance.q}) with increasing scale $s$. For the majority of cases considered herein, this increase was faster than the one of its counterpart; therefore, $d_{xy}(q)$ is usually positive. In~the cases of BTC/EUR and BTC/USD, both versus BTC/ETH, the opposite, however, applied for the smaller $q$-values. This signals that certain elements of the dynamics of cross-correlations in these two specific configurations are distinct.\par

A more global measure of the cross-correlations is based on the $q$-dependent detrended cross-correlation coefficient $\rho_q$ calculated according to Equation~(\ref{rhoq}). The time scale $s$-dependence of such coefficients, considered between the same exchange rate pairs as in Figure~\ref{fig:Fq-cc}, is shown in Figure~\ref{fig:rho-q} for $q=1$ (which filters out the medium size fluctuations) and for $q=4$ (large fluctuations). In~essence, all combinations of exchange rates are found to be correlated (the sign depends on whether an exchange rate or its reverse is taken), for both $q=1$ and $q=4$ (and similarly for the intermediate values of $q$), but they differ significantly from one another in the magnitude of $\rho_q$. The highest values correspond to BTC/EUR versus BTC/USD and ETH/EUR versus ETH/USD, with the correlation measure approaching almost unity at scales $s$ on the order of a few days. The lowest values, on~the other hand, correspond to BTC/EUR versus BTC/ETH and to BTC/USD versus BTC/ETH, both~saturating at around 0.2 for $q=1$ and at an even lower value for $q=4$. In general, however, in the present study, the degree of cross-correlation does not depend markedly on $q$; this is in contrast with mature stock markets and the Forex~\cite{kwapien2015,Zhao2017,Watorek2019}, wherein the cross-correlations viewed at $q=1$ are, typically, significantly stronger than those at $q=4$. This difference may originate from the fact that the amplitudes of the crypto-currency price fluctuations tend to be vastly larger than those hallmarking more traditional markets.\par
\begin{figure}[h!]
\centering 
\includegraphics[scale=0.5]{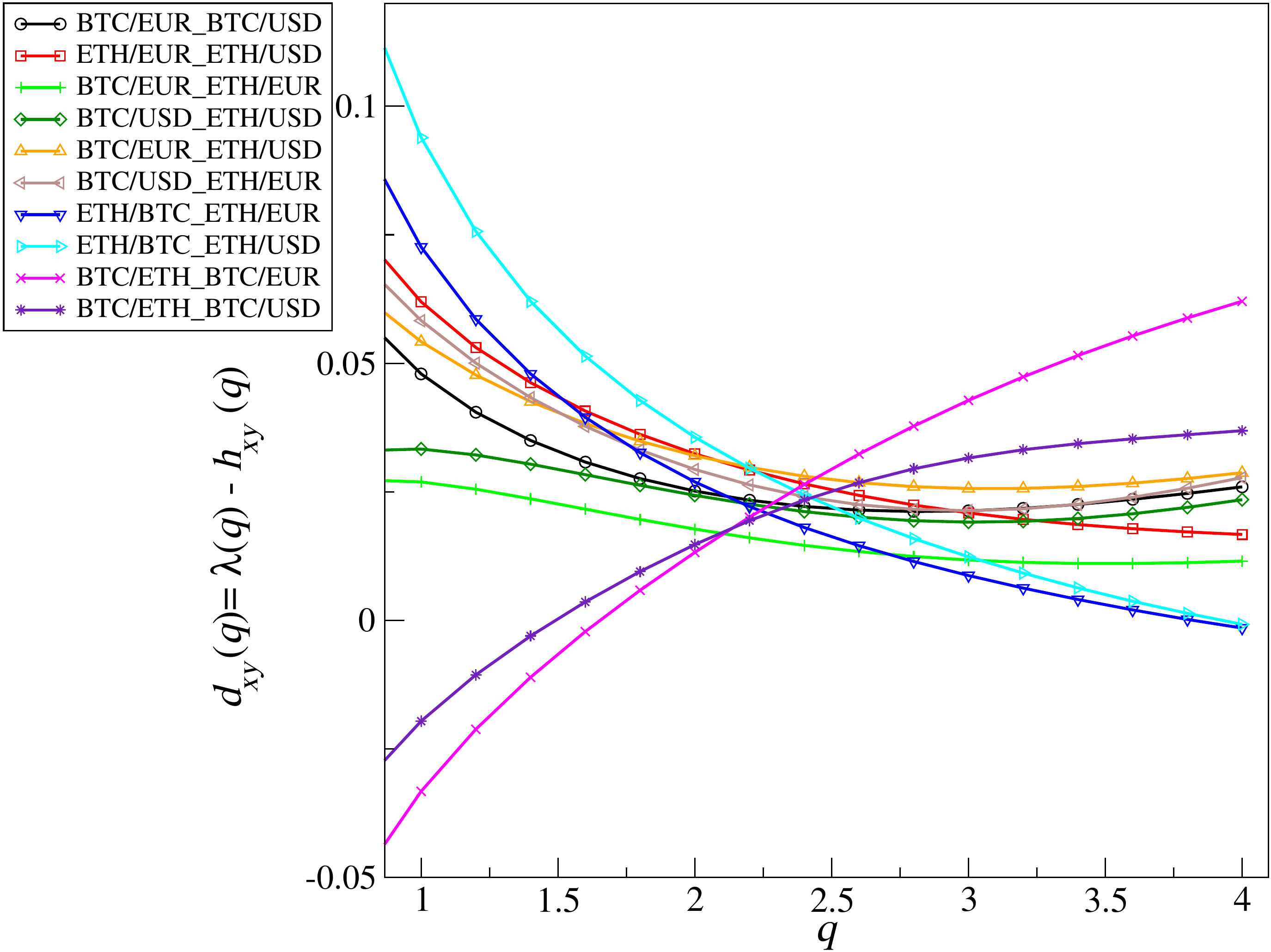}
\caption{(Color online) Differences between multifractal cross-correlation scaling exponents $\lambda(q)$ and the average generalized Hurst exponents $h_{xy}(q)$ estimated for $q\in [1,4]$ corresponding to the cases considered in Figure~\ref{fig:Fq-cc}.} 
\label{fig:lambda_h-average}
\end{figure}

A characteristic feature of the $\rho_q(s)$ coefficients in Figure~\ref{fig:Fq-cc} is their scale $s$-dependence, such that it reflects increasing correlation with decreasing frequency of probing price changes. Such effects are, in fact, complementary to the well-known Epps effect \citep{Epps,kwapien2004,Toth2009,drozdz2010} which appears in diverse financial contexts~\cite{Pal2014,Reboredo2014,kwapien2015,Li2016,Hussain2017,Zhao2017}. In the present case, a rather systematic correspondence between the scale $s$-dependence of $\rho_q(s)$ in Figure~\ref{fig:rho-q} and $d_{xy}(q)$ displayed in Figure~\ref{fig:lambda_h-average} can be traced. The larger $d_{xy}(q)$ is, the faster, on average, is the increase of $\rho_q(s)$ with $s$~\cite{Watorek2019}. When the fluctuation functions scale, which here to a good approximation holds true in most of the cases under consideration, this can be understood based on Equation~(\ref{rhoq}); $\lambda>h_{xy}$ implies that the numerator in this equation grows with $s$ faster than the denominator, and thus the effect is magnified with an increasing $d_{xy}$, which aligns with the observations in Figure~\ref{fig:rho-q}.\par

\begin{figure}[h!]
\centering 
\includegraphics[scale=0.5]{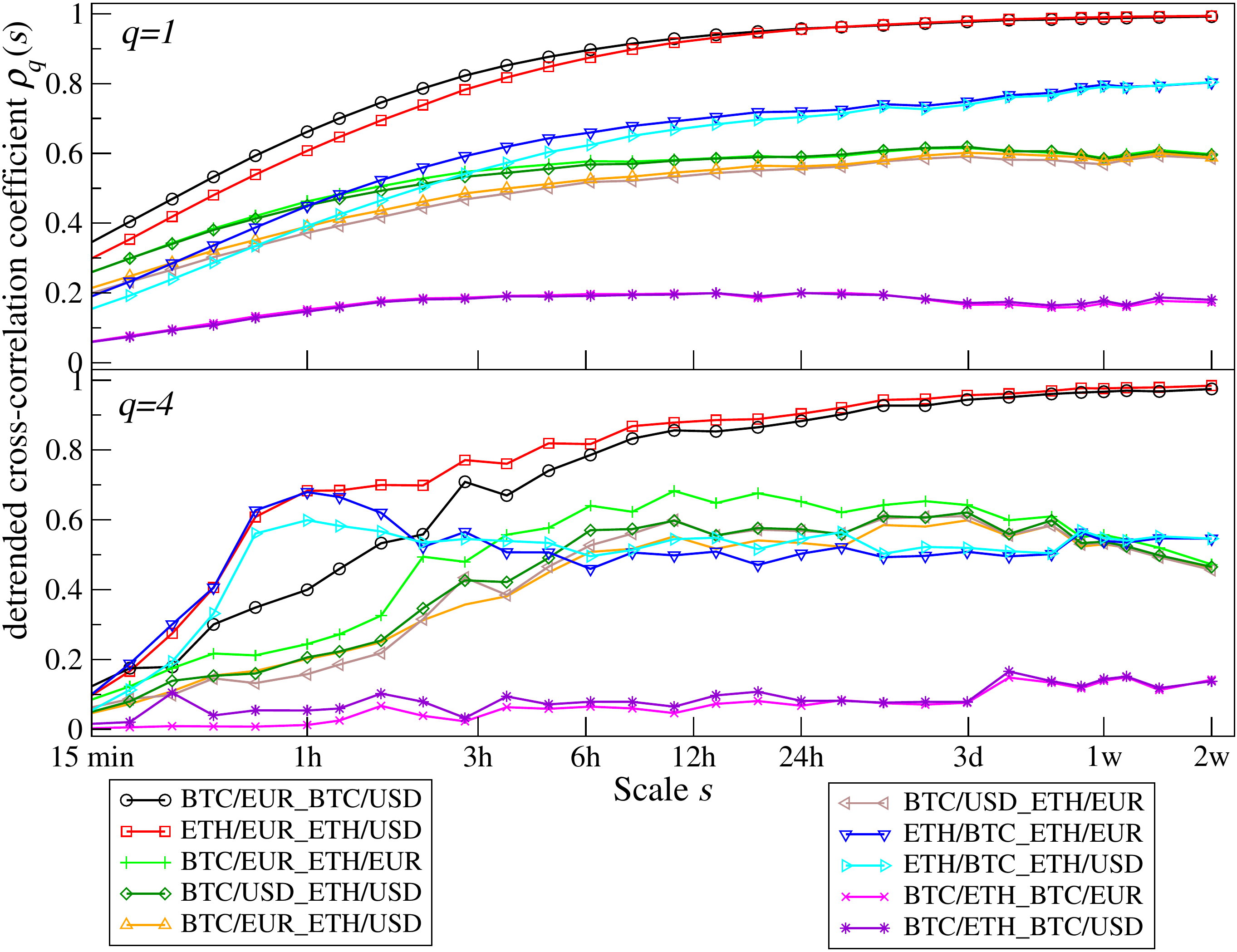}
\caption{(Color online) $q$-dependent detrended cross-correlation coefficients $\rho_q(s)$ for the same ensemble of exchange rates as in Figure~\ref{fig:Fq-cc}, shown as functions of the temporal scale $s$ for $q=1$ and $q=4$.}
\label{fig:rho-q}
\end{figure}

\section{The BTC/ETH versus the EUR/USD rates} 

In addition, Figure~\ref{fig:singspektrum} shows that, out of those exchange rates involving the BTC and/or the ETH, the most developed and least asymmetric singularity spectrum $f(\alpha)$ was found for the BTC/ETH rate, thus, for the one whereby the two crypto-currencies are exchanged directly between themselves. It~is in this sense that this particular exchange rate resembles the most, though of course not yet to a full extent, the EUR/USD rate. An important question that emerges in this regard, then, concerns~the cross-correlation between the BTC/ETH and the EUR/USD exchange rates. Since~both yield the broadest multifractal spectra, one may naively expect that their cross-correlation should also be markedly stronger than those considered in Figures~\ref{fig:Fq-cc} and~\ref{fig:rho-q}. The reason why this particular cross-correlation was not included in those figures in parallel with the other pairs of exchange rates shown there was that the EUR/USD is not traded during the entire week, whereas the BTC and the ETH are. The world Forex market, and thus also the EUR/USD, is traded from Sunday 10 p.m. UTC to Friday 10 p.m. UTC. For the particular purpose of assessing the degree of cross-correlations between the EUR/USD and BTC/ETH exchange rate changes, these breaks were thereafter removed, thus retaining in the time-series fully corresponding intervals for quantifying the cross-correlations.\par

Figure~\ref{fig:benek-vs-edek} shows the corresponding results for the fluctuation cross-covariance functions $F_{XY}(q,s)$ (upper panel) and the $\rho_q(s)$ coefficient (lower panel). Somewhat surprisingly in view of the previous results showing substantial cross-correlations seen in Figures~\ref{fig:Fq-cc} and~\ref{fig:rho-q} between all other exchange rates, the cross-correlation between BTC/ETH and EUR/USD is basically nonexistent. On the level of the $F_{XY}(q,s)$ functions, only at around the largest $q=4$ value one could see a trace of scaling and, down from $q \approx 2$, these functions start fluctuating between positive and negative values, thus~cannot even be shown in a log-log scale. An even more explicit negation of the existence of any significant cross-correlation between the BTC/ETH and the EUR/USD exchange rates is given by the $\rho_q(s)$ coefficient: its value remains close to zero for all $q$ and scale $s$ values.

\begin{figure}[h!]
\centering 
\includegraphics[scale=0.5]{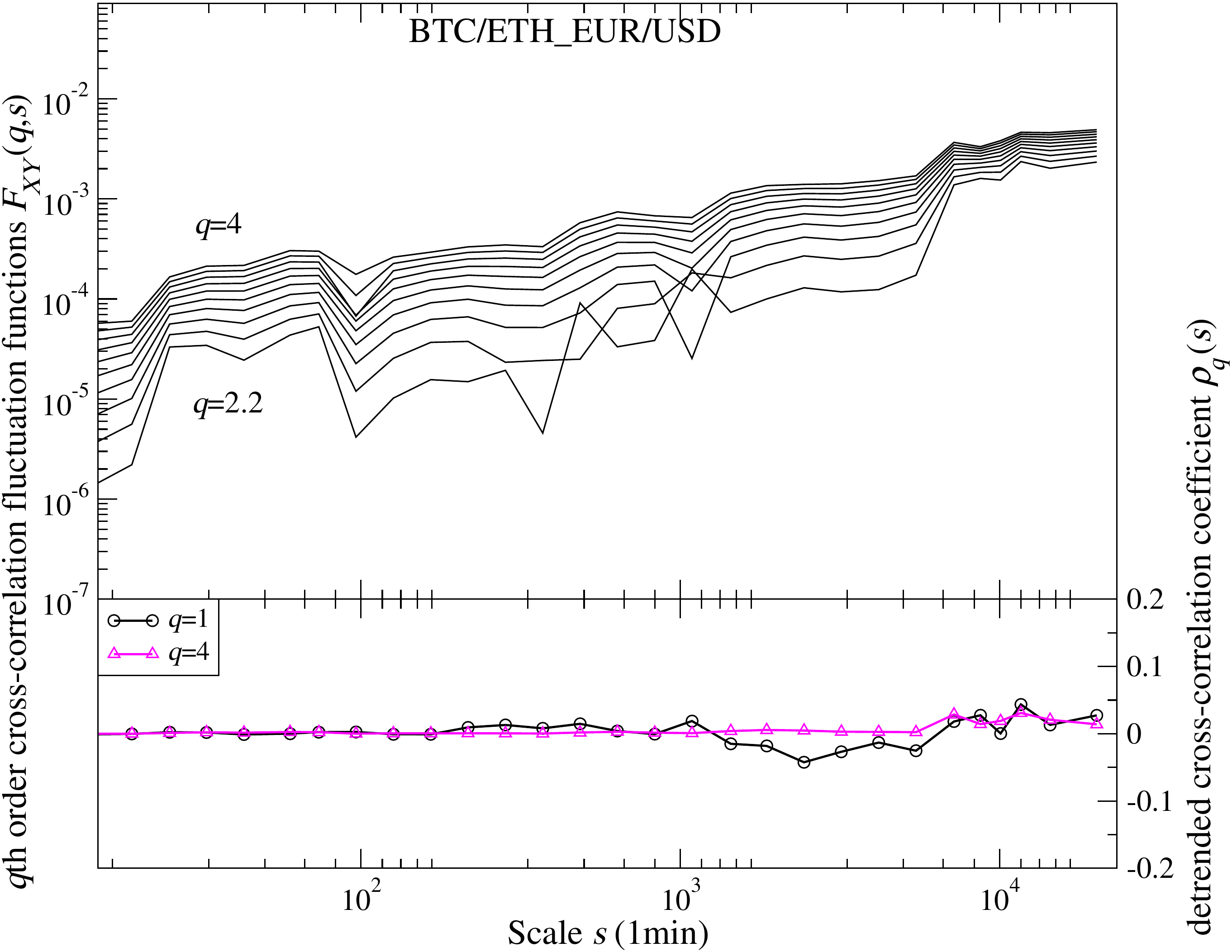}
\caption{(Color online) Family of the $q$th-order fluctuation cross-covariance functions $F_{XY}(q,s)$, for~different values of $q$ in steps of 0.2 The upper-most lines correspond to $q=4$, whereas the bottom ones to those for which $F_{XY}(q,s)$ are still positive (upper panel) and the $\rho_q(s)$ coefficients as functions of the temporal scale $s$ for $q=1$ and $q=4$, calculated for the cross-correlation between BTC/ETH and EUR/USD exchange rates (lower panel).}
\label{fig:benek-vs-edek}
\end{figure}

\section{Summary}
\label{summary}
As reported in a recent study~\cite{BTC2018}, in spite of its virtual nature and novelty, the Bitcoin (BTC) market over the years 2016--2017 developed the statistical hallmarks which are empirically observed in all 'mature' markets like stocks, commodities or the Forex. The present study not only confirms this observation for the year 2018 but also documents that it can be extended to another crypto-currency, the Ethereum (ETH), which, in terms of the capitalization involved can be considered the second most important one among the hundreds of crypto-currencies that nowadays circulate and are traded.\par Indeed, the fluctuations of the ETH exchange rates, with respect to both the EUR and the USD, to a good approximation obey the inverse cubic power-law. The autocorrelation functions for the cryptocurrencies have developed dynamics which resemble the EUR/USD rate just one decade ago}. Moreover, their Hurst exponent has been systematically increasing and approached the 'mature' value of 0.5 in early 2018. There is, furthermore, even evidence of a well-developed multifractality. Recently, the directly-traded exchange rates between the Bitcoin and Ethereum (BTC/ETH) appear to obey analogous fluctuation characteristics, with the multifractal spectrum being even broader and more symmetric compared to those of BTC or ETH in relation to fiat currencies like EUR and USD. There are also significant cross-correlations between the considered crypto/fiat exchange rates, similar to those observed in the mature financial markets~\cite{Watorek2019}. What, however, appears particularly thought-provoking is that, meanwhile, the cross-correlations between the BTC/ETH and the EUR/USD exchange rates entirely disappear.\par
Altogether, these facts, and especially the last one, provide strong support for the hypothesis~\cite{BTC2018} of the gradual emergence of a new and at least partially independent market, analogous to the global foreign exchange (Forex) market, wherein cryptocurrencies are traded in a self-contained manner. In~more practical terms, this means that not only the Bitcoin but even the whole emerging crypto-market may, eventually, offer 'a hedge or safe haven' for currencies~\cite{urquhart2019}, gold and commodities~\cite{shahzad2019}. \par
Analysis of the cross-correlations between cryptocurrencies also reveals time-scale and fluctuation-size dependence, which could be relevant in the design of trading strategies and portfolio management. In~the present study, the cross-correlations involving only two highest-capitalization cryptocurrencies are systematically explored on a fully quantitative level. Future work should explicitly address the larger set of most liquid cryptocurencies, also relevant to triangular arbitrage. Despite already featuring statistical characteristics close to the mature markets, the cryptocurrency market is still continuously evolving its nature; hence, future work should also address other periods and time-scales of its development.

\def\bibfont{\footnotesize}





\end{document}